# "Cryptographic Challenges: Masking Sensitive Data in Cyber Crimes through ASCII Art"

Andres Alejandre [1], Kassandra Delfin[1], Victor M Castano[2]

1.- IBM de México Campus Guadalajara.
Carretera a El Castillo 2220,
El Salto, Jalisco, 45680,
México.

2.- Centro de Física Aplicada y Tecnología Avanzada,
Universidad Nacional Autónoma de México
Boulevard Juriquilla No. 3001.
Santiago de Querétaro, Querétaro, 76230,
México.

Andres Alejandre (corresponding author: Andres.Alejandre@ibm.com)

Abstract

The use of ASCII art as a novel approach to masking sensitive information in cybercrime, focusing on its potential role in protecting personal data during the delivery process and beyond, is presented. By examining the unique properties of ASCII art and its historical context, this study discusses the advantages and limitations of employing this technique in various cybercrime scenarios. Additionally, providing recommendations for enhancing data security practices and fostering a culture of privacy awareness in both businesses and individuals. The findings suggest that ASCII art, with its simplicity and ambiguity, can serve as an effective tool against cybercriminals, emphasizing the need for robust data security measures and increased privacy awareness in today's interconnected world.

# Introduction

According to the Internet Crime Report Data by the Federal Bureau of Investigation, data breaches are one of the most common vulnerabilities attacks and just in the US (Federal Bureau of Investigation, 2023), in 2023, over 298 thousand individuals reported encountering phishing/vishing/smishing attacks. These incidents decreased by 0.5 percent as compared to the previous year, when the number of phishing attacks nationwide amounted to over 300 thousand (Federal Bureau of Investigation, 2023).

In the digital age, personal information protection has become a critical concern, leading tech giants like Apple, Meta, and Google to emphasize their commitments to privacy through robust security measures and transparent data practices. Despite these efforts, data breaches remain frequent, highlighting a gap between public assurances and actual data security. Research shows that even companies which heavily invested in cybersecurity can fall short, exposing the ongoing challenges in protecting user data in an ever-evolving threat landscape (Martin & Shorey, 2023).

The delivery of goods to consumers has become an indispensable component of contemporary e-commerce. Companies like Amazon rely on services such as DHL and UPS to ensure that products reach customers efficiently and, in principle, safely. However, the process of delivering packages involves the collection and use of personal information, including names, addresses, and sometimes even contact details. While this data is crucial for ensuring the package reaches the correct location, it raises significant privacy concerns. For instance, once a package is delivered, the information about the recipient's location and identity remains vulnerable. This can occur in several ways: the delivery person might remember the details, packages could be stolen, or discarded shipping labels could end up in the trash, potentially exposing sensitive information to anyone who comes across them. These risks underscore the importance of stringent data handling practices even after the initial delivery is completed, as the protection of personal information does not end with the receipt of a package (Woo, Jang, Woojung, & Hyoungshick, 2020).

ASCII (American Standard Code for Information Interchange) is a character encoding standard developed in 1963 that comprises 128 characters, including control characters, letters, digits, and symbols, standardized under ANSI X3.4 (Kelechava, 2019). This code became a fundamental component of early computing, facilitating consistent text representation across different systems and enabling the creation of text-based visuals. One notable application is ASCII art, a graphic design technique that constructs images using the 95 printable characters defined by the original ASCII standard, as well as extended proprietary character sets beyond the standard 7-bit code, which gained popularity on pre-internet platforms like bulletin board systems (BBSs) and Telnet (Kelechava, 2019).

E-commerce delivery companies play a crucial role in determining when and how a customer receives their order. The carrier's responsibility begins with transferring an order from the online store and concludes with delivering it to the customer's doorstep (Shyrma, 2024). Due to this significant role, it is essential for companies to implement safer delivery methods to protect their clients' personal information and ensure secure package delivery. With rising global demand for home deliveries, as seen with 97.6% of consumers in Asia and Latin America opting for this method, secure, reliable delivery solutions are more important than ever. Additionally, the emergence of recollection points, particularly in regions like Oceania where 12.2% of shoppers favor this option, highlights the need for flexibility in delivery approaches that cater to both customer preferences and security concerns (Myint Soe, 2023).

Using ASCII Art to mask the client's personal information ensures that sensitive details, such as their name and address, are hidden from unauthorized view. Rather than displaying the information in plain text, it is converted into a visually encoded image that can only be deciphered by individuals with the correct tools, such as the delivery personnel. This method enhances privacy by restricting access to essential data, ensuring that only authorized personnel can decode the information to deliver the package to the correct address; which brings the question: how much do you really value your personal information and security?

Methods and Procedures

To analyze the different patterns of an ASCII art code to decode the encrypted information, we propose here a two-step process. In the first step, the sensitive data is encrypted and injected into the generated ASCII art image, replacing some of the hash that was initially there. The encryption process involves generating a unique key for each data set and applying a hash function to it. This ensures that the encrypted data remains unreadable without the corresponding decryption key (Alejandre, Llamas, Orozco, & Orozco, 2024), as schematically shown in figure 1.

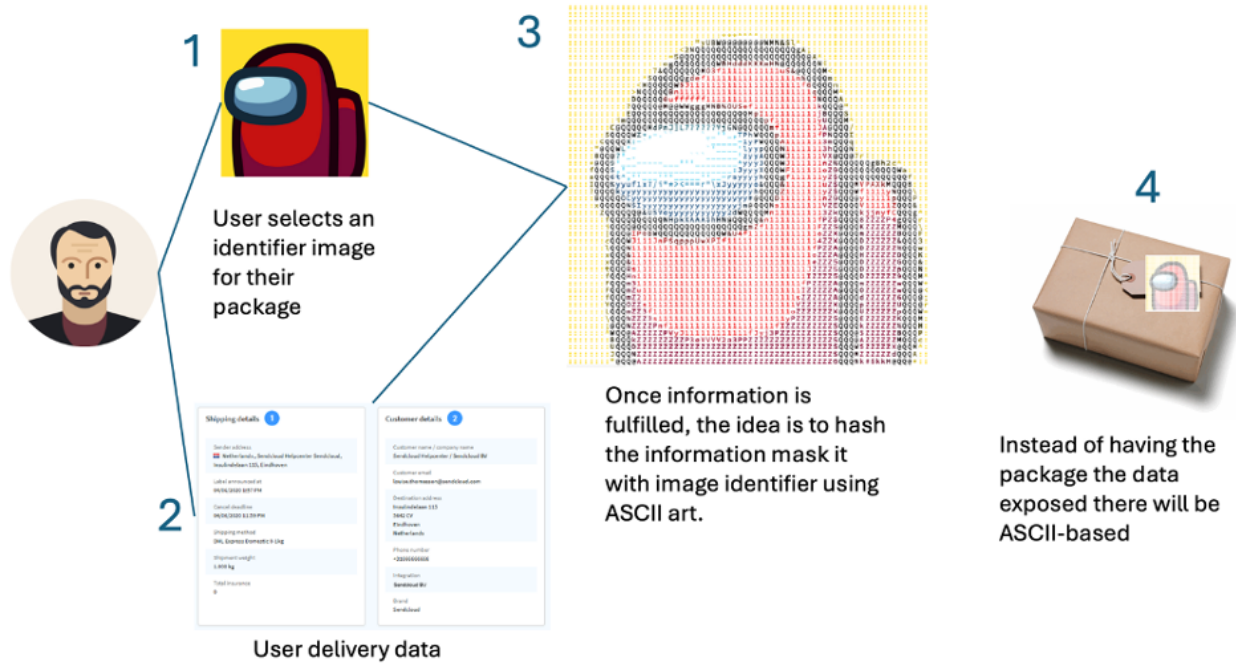

*Source: Own Creation.*

In the second step, the user attempts to decrypt the information by selecting specific areas within the ASCII art image. These areas serve as reference points or coordinates for the decryption process. Once the reference points are identified, the user can utilize the unique key generated in the first step to decrypt the encrypted data and reveal the sensitive information.

To further enhance security, an optional third layer can be implemented. This layer involves incorporating color into the decryption process. By checking whether the color of the selected area matches the expected color (defined by a HEX color code), the system adds an additional layer of complexity to the decryption process. This makes it even more difficult for unauthorized users to decipher the information without the correct color match.

Throughout the process, the system must validate that the image has not been significantly modified, as ASCII images are highly susceptible to alterations. This validation step ensures that the decrypted information remains accurate and unaltered.

The system should be implemented end-to-end, with clear distinctions between the responsibilities of the main company (creating the ASCII image with encrypted information) and the users (decrypting the information using the reference points provided). This ensures that confidential information is handled securely and only accessible to authorized individuals.

The diagram of figure 2 illustrates the first step, in this case we are not considering the ACSII image creation, the user needs to set or define the areas in the ACSII image that will have the information encrypted, after that the system receives the information and encrypts it, the last step is to inject the encrypted information in the previously selected areas.

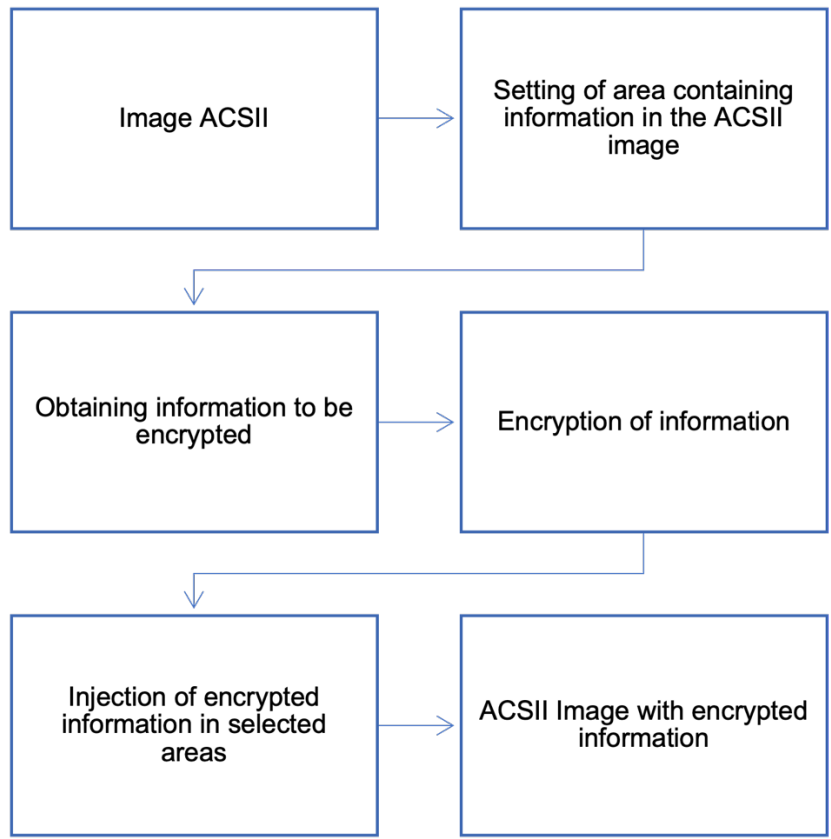

*Source: Own Creation.*

The diagram of figure 3 describes the consecutive steps when the user tries to use the encrypted information in the ACSII image. To decrypt the information, the user only needs to have the reference points or coordinates, using this data the user will be able to decrypt the information contained in the ACSII image.

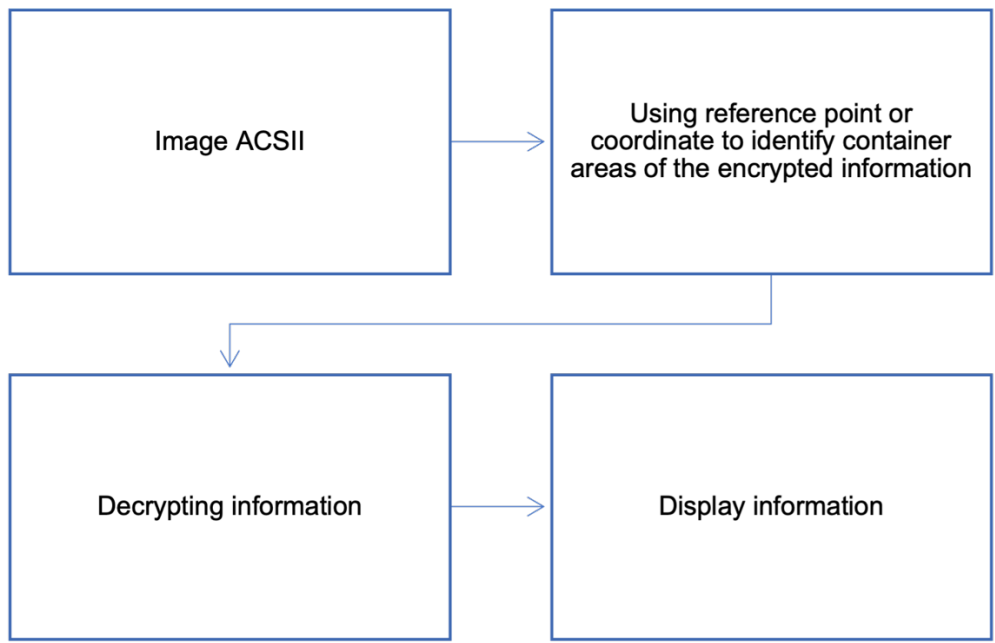

*Source: Own Creation.*

Results

To demonstrate the practicality of this approach, consider a hypothetical delivery service scenario. In this case, a user could choose to encode their delivery information—such as address, name, and references—within an ASCII art image. Upon delivery, the recipients would only be able to view the information if they possess the correct decryption key, which would be provided to them upon request. This ensures that the sensitive information remains confidential, and it is only accessible to authorized individuals.

Figure 4 illustrates how the ASCII art image works, showcasing different layers and the ability to hide encoded messages within the image. Only those with the correct decoder can access the information, while others will see only a simple image with no valuable information on sight, as illustrated in figure 4.

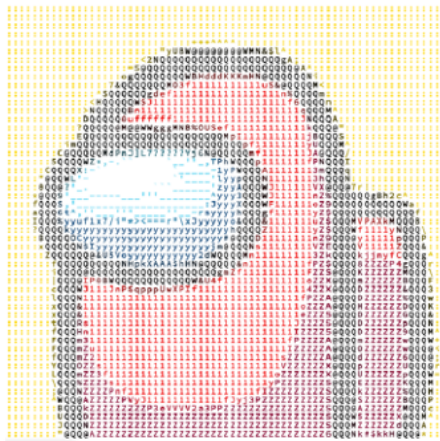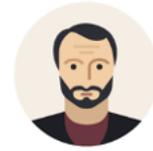

*Source: Own Creation.*

This would be particularly useful for scenarios where users want to protect their personal information, such as delivery services, online transactions, or any other situation where sensitive data needs to be shared securely.

To further evaluate the effectiveness of our method, we conducted a series of experiments. In these tests, we successfully decoded encrypted information embedded within ASCII art images, demonstrating the viability of our solution. We also assessed the robustness of our system by evaluating its resilience to common image editing tools and techniques. Our results indicate that the ASCII art encryption and decryption method is a reliable and secure approach for protecting sensitive information.

Discussion

This research highlights the potential of ASCII art encryption and decryption as a powerful tool for safeguarding sensitive data by combining the visual appeal of ASCII art with the security of encryption techniques, so a robust and user-friendly system for protecting confidential information, can be constructed. Further investigation into the application of this method to various industries and scenarios is warranted, as we believe it has the potential to significantly impact the way sensitive information is handled and protected.

Additionally, we have explored the integration of encryption, steganography, and secret splitting techniques to create a more complex and secure secret sequenced message based on ASCII art. Instead of embedding encryption/steganography metadata as part of the secret message, we propose sharing such metadata via offline means. This approach further enhances the security and complexity of the secret message, making it more resistant to unauthorized access and tampering.

The users in the communication of the encrypted data will define the information search parameters in the ASCII image, providing reference points, coordinates or any other element necessary for the location of the information. The definition of the information location in the ASCII art, will not be limited only in a particular sector, but also several locations of the image can be defined, in a certain defined sequence, so that later they can be concatenated or used separately, depending on the needs.

This idea may be used by any company or group of people who must handle visible information, for example those in charge of distributing packages, such as UPS, FEDEX, EXPRESS, etc., which contain sensitive or confidential information in the packages, such as name, address, phone, etc. In this case, instead of displaying this explicit information, the package will have only the ASCII image, which will contain the encrypted data in different locations and camouflaged with the existing ASCII code.

This technique, along with other forms of encoding, can serve as a means of safeguarding sensitive data by transforming it into a less obvious format. While effective in certain contexts, these methods also highlight the growing complexity of data protection in the digital age, where the line between security and accessibility must be carefully managed.

As mentioned, the use cases involve converting an image into ASCII art and using it as the information, can be posted in a delivery package to ensure encrypted information and enable data retrieval.

- Secure package identification: By converting the image into ASCII art and embedding encrypted information within the characters, each package can have a unique identifier. This identifier can be decrypted to obtain information about the recipient or the package itself. It provides a secure and visually appealing way to label packages with encrypted data.
- Efficient package tracking: The code embedded into ASCII art can contain encoded data that represents package tracking information. The information can be split into different parts into the images as described and that information could be read based on that. The purpose is also to make sure the information is distributed but also that it can be used with multiple packages to decrypt the info.
- Secure Messaging: The method could be applied to encrypt and hide sensitive information for messaging platforms. For instance, this method could be used on slack or email messaging for sending information which could not be human readable, but as soon as the endpoints share the private key, the information can be decrypted as soon as one of the endpoints use the device and take a photo for decrypting the information.
- Digital Watermarking: By embedding hidden information within the ASCII art, such as ownership details or copyright information, the art can serve as a unique identifier or proof of authenticity for digital assets.

## Conclusion

The proposed solution, which involves the use of ASCII art encryption and decryption, offers a promising approach to protecting sensitive information. By analyzing the patterns of an ASCII art code, we can decode encrypted information and reveal the underlying sensitive data. This method provides an additional layer of security beyond traditional data obfuscation techniques, as the encoded information is disguised within a visually appealing ASCII art image.

The proposed ASCII art sequential method, when combined with other encoding techniques, offers a promising solution for safeguarding sensitive information. As the complexity of data protection continues to evolve, it is crucial for organizations to stay informed about emerging encoding methods and their potential applications. By doing so, they can develop effective strategies for managing sensitive data while ensuring accessibility for intended recipients.

This new approach for cyber security, when partnering with the right company, can impact many people across multiples countries as per our example of delivery services. This can be applied to digital marketplaces or even postal services, making this a big player in society where the personal/sensitive information can be better protected for the citizens.

Finally, Cyber security is a very big topic nowadays, but we still need to keep in mind that our information is not only made available in a digital environment, but also in a physical manner in many different places and extreme care should be taken, such as we would for our online information.